# Ion heating in the PISCES-RF liquid-cooled high-power, steady-state, helicon plasma device

S. Chakraborty Thakur,[1, 2, a), b)] M. Paul,[3, b)] E. M. Hollmann[1], E. Lister,[4] E. E. Scime,[3] S. Sadhu,[5] T. E. Steinberger,[3] and G. R. Tynan[1]

[1] *Center for Energy Research, University of California at San Diego, San Diego, CA 92093*
[2] *Department of Physics, Auburn University, Auburn, AL 36849*
[3] *Department of Physics and Astronomy, West Virginia University, Morgantown, WV 26506*
[4] *Department of Mechanical Engineering, Carnegie Mellon University, Pittsburgh, PA 15213*
[5] *Department of Mathematics, Georgia College & State University, Milledgeville, GA 31061*

Radio Frequency (RF) driven helicon plasma sources are commonly used for their ability to produce high-density argon plasmas ($n > 10^{19}$ m$^{-3}$) at relatively moderate powers (typical RF power < 2 kW). Typical electron temperatures are < 10 eV and typical ion temperatures are < 0.6 eV. A newly designed helicon antenna assembly (with concentric, double-layered, fully liquid-cooled RF-transparent windows) operates in steady-state at RF powers up to 10 kW. We report on the dependence of argon plasma density, electron temperature and ion temperature on RF power. At 10 kW, ion temperatures > 2 eV in argon plasmas are measured with laser induced fluorescence, which is consistent with a simple volume averaged 0-D power balance model. 1-D Monte Carlo simulations of the neutral density profile for these plasma conditions show strong neutral depletion near the core and predict neutral temperatures well above room temperatures. The plasmas created in this high-power helicon source (when light ions are employed) are ideally suited for fusion divertor plasma-material interaction studies and negative ion production for neutral beams.

## 1. Introduction

Radio frequency (RF) based plasma sources are extremely popular in basic plasma physics experiments because of their ability to produce relatively high-density plasma at relatively low RF power inputs [1 – 5 and the references therein]. Depending on the source parameters (external magnetic field, gas pressure, gas flow rate and RF power input), these RF devices operate in the capacitive (E), inductive (H) or the helicon mode (W). In the helicon mode of operation, RF sources produce centrally peaked plasmas with densities (*n*) of ~ $10^{19}$ m$^{-3}$ in argon for a few kW (kilowatts) of RF power in both small volume semiconductor etchers and large volume plasma chambers that are several meters long [6, 7, 8].

Helicon plasma sources have the potential to be used in plasma material interaction (PMI) devices that simulate fusion plasma divertor-like conditions [9 – 12], if high densities can be achieved in steady-state conditions in fusion-relevant gases such as deuterium (D), hydrogen (H), and helium (He). Helicon plasma sources have several advantages over existing PMI devices, typically based on reflex arc sources or heated cathode sources [12] that can introduce significant levels of impurities into the plasma. By mounting the RF antenna outside of the vacuum chamber, it is anticipated that impurities from the source can be minimized. For RF sources, axial access to the vacuum chamber is unobstructed by the source itself (unlike the arc discharge sources) and is available for laser injection and diagnostic access. Another important feature of an RF source is that they generate very repeatable plasmas even after long device downtimes.

The plasma density in these linear RF devices is determined by balancing the plasma production against radial turbulent losses together with plasma flow speeds that are proportional to the ion thermal speed along the magnetic field, resulting in enhanced parallel end-plate losses for light ions. As a result, the high thermal velocities of light ions in RF plasmas necessitates RF powers of order of 10 kW or more to achieve fusion divertor-like plasmas densities. Hence there are ongoing research efforts to build high-power RF sources for PMI studies [13 – 15]. A challenge in using helicon plasma sources for PMI studies is that the electron and ion temperatures in helicon plasmas are relatively low (typical electron temperature, $T_e$ ~ 5 eV and typical ion temperature, $T_i$ < 0.7 eV [16 – 18]). The low temperatures are ideal for probe diagnostics in basic plasma studies (probes can be inserted across the entire plasma cross section without getting damaged or perturbing the plasma). However, reproducing fusion plasma divertor conditions requires higher electron temperatures and, more importantly, higher ion temperatures. In addition, achieving steady state conditions is important to generate the high fluences necessary to perform longtime material exposure studies.

In the literature, the highest ion temperature reported in high-density, steady-state helicon sources is ~ 0.7 eV [16 – 18, 20 – 23], for magnetic field strengths of ~ 1000 G and a maximum RF power of 4 kW [23], though typically helicon sources operate at RF powers less than 2 kW. Higher ion temperatures, over 2 eV, have recently been observed in low-density plasma plumes as a helicon source plasma expands into a weaker magnetic field [24]. The 4 kW study involved a source, RAID (Resonant Antenna Ion Device) [25], being developed

---

for negative ion production for neutral beams and if those results are scaled to an RF power of 10 kW, ion temperatures greater than 1 eV are predicted [23].

Steady-state helicon operation at RF powers of ~ 10 kW poses several challenges. A helicon plasma source requires a cylindrical dielectric window (an insulating sleeve) for the RF to penetrate into the plasma. Thermal loading of the dielectric window (see Fig 1) limits existing high-power helicon plasma sources used for PMI studies to pulsed operation with maximum pulse lengths of hundreds of milliseconds [11, 13 – 15]. To solve the thermal loading issue while maintaining the vacuum integrity of the RF source, we have designed and built a novel liquid-cooled RF window for steady-state operation with up to 20 kW of RF power on the PISCES-RF facility [26]. De-ionized (DI) water, chosen to minimize RF absorption, flows between two concentric cylindrical insulators to remove the heat. The water is cooled with a dedicated recirculating chiller. This is the first time an RF-transparent dielectric window has been completely immersed in a full azimuthal blanket of liquid coolant. This design ensures uniform cooling. The PISCES-RF design is a prototype for the next generation PMI device, Materials Plasma Exposure eXperiment (MPEX), under construction at the Oak Ridge National Laboratory (ORNL) that is planned to operate at up to 200 kW of steady-state RF power [27, 28]. The conceptual design and preliminary numerical analysis of a water-cooled ceramic window for the MPEX device is based on thermal data and numerical simulations of the thermal-mechanical stresses [29, 30] for the short pulse, uncooled source in use on the Proto-MPEX facility [11]. In a companion paper, we describe in detail the effects of DI-water coolant on plasma production, RF absorption by the coolant, thermal infra-red imaging of the liquid-cooled ceramic assembly and hydrogen plasma production in PISCES-RF [26].

Here we report the performance of the DI-water-cooled RF plasma source for argon plasmas at up to 10 kW of steady-state operation. Although argon is used here, many results of this work, such as the importance of neutral depletion that determines the plasma density limit, and the effectiveness of using a water cooled RF window for the source, are also relevant for hydrogen and helium plasmas. The measurements include radial profiles of the plasma density and electron temperature and the core ion temperatures obtained through non-perturbative laser induced fluorescence (LIF). LIF is routinely employed in helicon plasma sources to measure Doppler-resolved velocity distribution functions (VDF) of argon ions, argon neutrals, helium neutrals, and xenon ions [31 – 33]. Achieving steady state plasma operation using the DI-water cooled helicon antenna enabled long integration times for these LIF measurements even for such high RF powers, up to 10 kW.

With the experimental profiles as input parameters, we also show results from a 1-D Monte Carlo simulation of the radial profile of the argon neutral density. The model predicts strong neutral depletion near the core of the plasma and significant charge exchange coupling of the neutrals with the ions. Understanding the details of the neutral density distribution is crucial to predicting the performance of PMI facilities like MPEX [34].

In addition to PMI studies, there are other basic science motivations to develop high ion temperature RF plasma sources. In previous work at relatively lower RF powers (< 750 W), the perpendicular ion temperature was found to increase with increasing magnetic field, while the parallel ion temperature remained relatively constant [16 – 18]. This ion temperature anisotropy (ratio of the perpendicular to the parallel ion temperature) was used to drive electromagnetic instabilities similar to those observed in space [18]. High ion temperature helicon plasmas at higher RF powers are expected to extend studies of plasma instabilities driven by ion temperature anisotropy to more space-relevant conditions. In the solar wind and the magnetosphere, these instabilities (the mirror and firehose instabilities) play a critical role in energy transport by limiting the maximum values of ion thermal anisotropy [19].

In section 2, we describe the experimental setup including details of the novel liquid-cooled helicon antenna assembly and the experimental device. In section 3, we describe the diagnostics, including a new, compact, portable, diode-laser-based LIF system [35]. In section 4 and 5, we present the experimental results and neutral modelling studies, respectively. In the appendix, we describe a deconvolution method developed to correct for artificial broadening of VDF measurements due to oversampling and a simple volume averaged 0-D power balance model to estimate ion temperatures.

## 2. Experimental set up

### 2.1 DI-water cooled RF source assembly

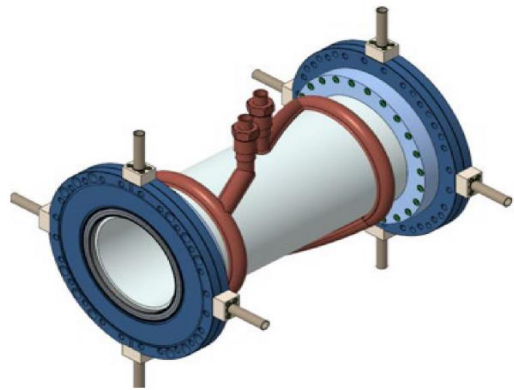

Fig. 1. CAD model showing the DI-water cooled helicon antenna assembly design.

A CAD (computer-aided design) model of the liquid-cooled, RF antenna assembly is shown in Fig. 1. The RF-transparent window (shown as the white cylinder), is comprised of two ceramic cylinders and the DI-water layer flows between them. The inner ceramic is an alumina cylinder ($Al_2O_3$) of thickness ~ 6.35 mm (0.25 inches) and inner diameter of ~ 12 cm (4.75 inches). The outer window is a concentric quartz cylinder of thickness

~ 3 mm (0.117 inches). The annular DI-water channel between the two tubes is ~ 3.35 mm (0.133 inches) wide. The RF-transparent ceramic window assembly thus has an outer diameter of ~ 14.6 mm (5.75 inches). The helical RF antenna (shown in reddish-brown in Fig. 1) is made from copper pipes and is independently water cooled. The DI-water is introduced via stainless steel tubing at four azimuthal locations (shown in grey) via the Conflat™ flanges (shown in blue) on the two ends. Fluid dynamics and thermal-hydraulic simulations show that the DI-water effectively removes the heat from the RF-transparent ceramic assembly [29, 30] and have been subsequently experimentally verified [26].

## 2.2 The plasma device

Shown in Fig. 2 is a CAD model of the PISCES-RF experimental device. The 3 m long, 20 cm diameter cylindrical, stainless steel plasma chamber, and the external magnets and pumping systems are reused from the Controlled Shear Decorrelation eXperiment (CSDX) [8, 36 – 46]. Previously, in CSDX, argon plasmas with high density and relatively low temperatures ($n > 10^{19}$ m$^{-3}$, $T_e$ ~ 5 eV and $T_i < 0.7$ eV [17, 19]) were produced using an $m = 1$ helicon antenna operating at 13.56 MHz at RF powers of ~ 1.5 – 2 kW. The new RF source assembly with the DI-water-cooled RF-transparent window (Fig. 1) will be used primarily for PMI applications – PISCES-RF (Plasma Interaction Surface Component Experimental Station – Radio Frequency).

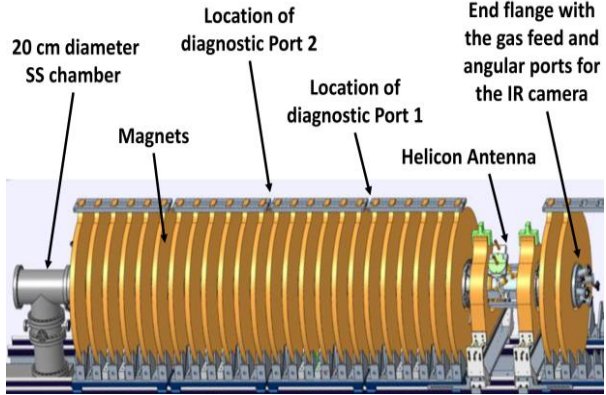

Fig. 2. CAD model of the PISCES-RF device.

The chamber is immersed in an external magnetic field of up to 2400 Gauss. A turbomolecular pump, backed by two mechanical pumps, at the downstream end of the chamber, maintains a base pressure in the high $10^{-7}$ Torr range. A manually controlled valve placed in front of the pumps at the end of the device facilitates changes in the neutral pressure independent of the gas flow rate. A specially designed flange placed upstream of the RF source enables upstream gas injection. This gas injection configuration has been shown to be the most effective for neutral gas management and operation of the Proto-MPEX source [34]. There are two primary locations for diagnostics, labeled Port 1 and Port 2 in Fig 2. The diagnostics ports are 0.8 and 1.5 m away from the RF source assembly, respectively.

## 3. Diagnostics

### 3.1 Laser Induced Fluorescence

For LIF [31 – 33] measurements of the ion temperature and the ion velocity, the three-energy level argon LIF scheme shown in Fig. 3 is used. The initial state is pumped with a tunable diode laser with central wavelength of 668.6138 nm; corresponding to the $3d^4F_{7/2}$ to the $4p^4D_{5/2}$ level transition in Ar II. The fluorescence emission at 442.7244 nm from the decay of the $4p^4D_{5/2}$ state to the $4s^4P_{3/2}$ is recorded as a function of laser wavelength. The transition includes both circularly polarized $\sigma$ ($m = \pm 1$) and linearly polarized $\pi$ ($m = 0$) transitions. The laser is injected perpendicular to the magnetic field with its polarization axis aligned with the magnetic field. Thus, only the $m = 0$ absorption lines are excited. For a magnetic field strength of 1000 Gauss or less, the effect of the Zeeman splitting of the individual $\pi$ lines are much smaller than the Doppler broadening and is ignorable in the analysis [47, 48]. Previous LIF measurements on CSDX, at lower RF power (1.5 kW), but similar magnetic fields have shown the validity of this approach [20, 22, 49 – 52]. The diode laser has a line width of ~ 1 MHz and can be scanned over a 30 GHz mode-hop-free tuning range, thereby allowing the measurement of the ion VDF (IVDF). To determine the perpendicular ion temperature and velocity, a nonlinear least squares fitting routine fits the measured IVDF with a drifting Maxwellian distribution of the form

$$I_R(v) = I_R(v_o) e^{\frac{-m_i c^2 (v-v_o)^2}{2kT_i v_o^2}}, \qquad (1)$$

where $v_o$ is the rest frame frequency of the absorption line, $m_i$ is the ion mass, and $T_i$ is the ion temperature.

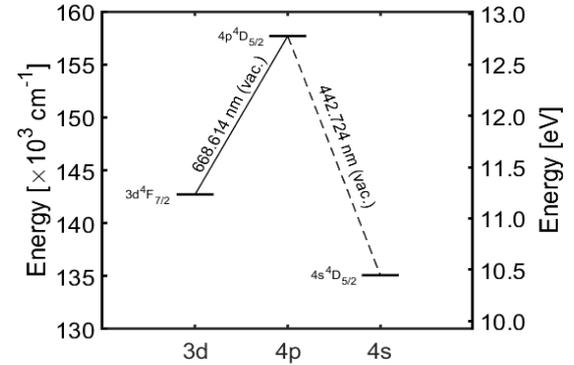

Fig. 3. Three-energy level LIF scheme used in this study.

### 3.2 LIF apparatus

The portable LIF apparatus is shown schematically in Fig. 4. The laser is a Toptica TA 100 tunable diode laser consisting of a Littrow configuration master oscillator followed by a tapered amplifier [35] capable of generating ~ 250 mW of output laser power. A small fraction of the original laser light from the master oscillator (~ 3 mW) is coupled into a 5-micron diameter single mode optical fiber and coupled to a Burleigh WA – 1500 wavelength meter, for real-time wavelength monitoring. The wavelength was measured to an accuracy of ± 0.0002 nm. The error in the measured velocity due to this instrumental error is ± 100 m/sec. Instead of an external chopper, the current driving the tapered amplifier is modulated by the internal reference frequency from an SR830 lock-in-amplifier. The output laser emission is coupled into a 200-micron diameter multimode optical fiber and transported to the plasma chamber via the laser injection optics. The laser injection optics consist of a collimating lens and a linear polarizer to eliminate the $\sigma$ transition components.

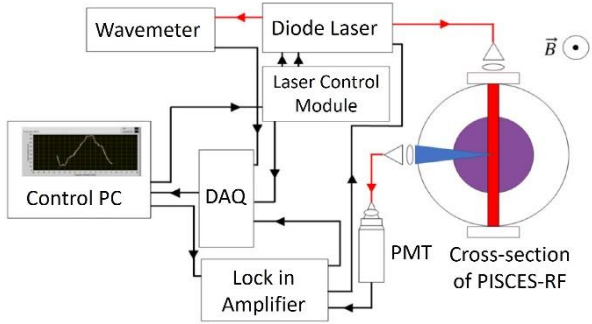

Fig. 4. Schematic of the LIF apparatus in PISCES-RF.

The fluorescent emission was collected by another lens, collimated, and then coupled into another optical fiber. The light exiting the fiber was re-collimated, passed through a 2 nm wide filter centered at 443 nm to reduce spontaneous emission signal, and measured with a visible light Hamamatsu 10493-011 photo-multiplier tube (PMT). The intensity of the fluorescent emission from the excited state as a function of laser frequency gives a direct measurement of the IVDF in the region where the injected beam overlaps the collection volume, resulting in a spatial resolution of ~ 3 mm. Since the PMT signal was composed of background spectral radiation, electron-impact-induced fluorescence radiation and electronic noise, the Stanford Research SR830 lock-in amplifier was used for phase synchronous detection of the LIF signal. Lock-in amplification was essential since electron-impact-induced emission was several orders of magnitude larger than the fluorescence signal.

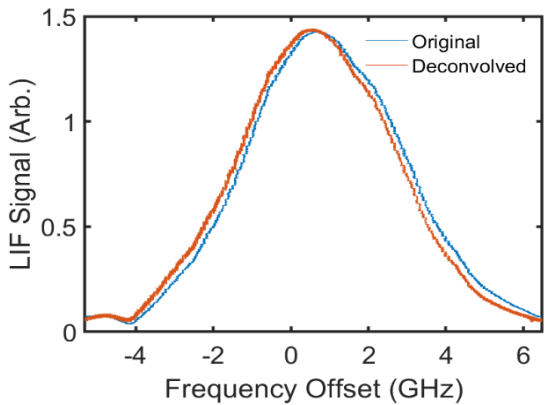

Fig. 5. Comparison between the original LIF signal and the deconvolved signal. For each case, the temperature is the same to within fitting error (0.72 eV). Before the correction, the measured flow is ~ 580 m/s. After the correction, the measured flow is ~ 440 m/s.

If the ratio of the total laser scan time ($T_{scan}$) and the lock-in time constant ($RC$) is small, there can be systematic errors leading to artificial broadening of the IVDF. A deconvolution technique (detailed explanation given in Appendix I) is used to rectify any artificial broadening due to this effect. For the LIF measurements shown in this work, the $T_{scan}/RC$ ratio was ~ 33 (100 sec scans with 30 sec integration time of the lock-in amplifier). The result of this deconvolution technique on an experimentally measured IVDF is shown in Fig. 5. For this example, there is little difference between the original and the reconstructed signal. Both yield the same fitted ion temperature, but the deconvolved IVDF yields a smaller bulk flow (shift of the distribution). The effects of the deconvolution technique are more dramatic for smaller values of $T_{scan}/RC$.

### 3.3 RF compensated Langmuir probes

We used two RF-compensated Langmuir probes [53] to measure the plasma density and electron temperature at two axial locations, 0.8 m and 1.5 m downstream of the RF antenna. A 10 nF capacitor and a series of five inductors (RF chokes) were used to effectively block the 13.56 MHz source frequency and its two lower and upper harmonics from affecting the probe characteristics.

## 4. Experimental Results

### 4.1 Confirming the helicon mode of operation

The classic high-density helicon mode of operation (W) is achieved by increasing either the RF power or the external magnetic field. The transition to the helicon mode is indicated by a sharp discrete jump in the core plasma density (along with changes in visible plasma emission and formation of centrally peaked radial profiles) [1, 54, 55]. The plasma density measurements at $r = 0$ confirming helicon mode of operation (W) are shown in Fig. 6 (increasing RF power) and Fig. 7 (increasing magnetic field).

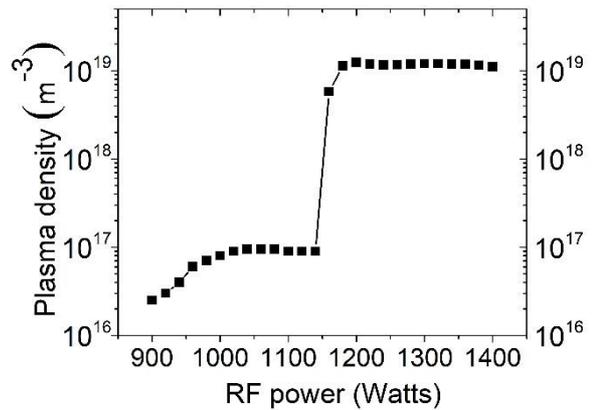

Fig. 6. Plasma density measured at $r = 0$, as a function of forward RF power. The sharp increase at 1160 Watts represent the transition to the helicon mode (at 320 A of current in the magnet coils, corresponding to ~ 1000 G).

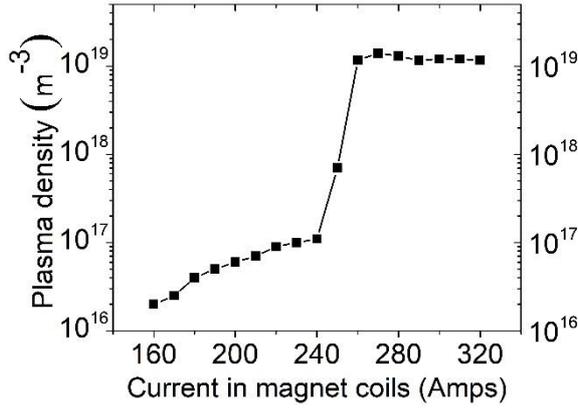

Fig. 7. Plasma density measured at $r = 0$, as a function of increasing current in the magnet coils. The sharp increase at 260 A (~ 800 G) represent the transition to the helicon mode (at 1.5 kW of forward RF power).

These measurements demonstrate that the DI-water coolant does not hinder high-density, steady-state plasma production, even in the helicon mode of operation. All results in this work are from steady-state operation, with the chilled DI-water effectively cooling the RF antenna assembly. Based on this experience, this same source and device was used to produce high density hydrogen and helium plasmas at 20 kW of RF power [26].

**4.2 Plasma density and electron temperature profile measurements: RF power scan**

For the RF power scan, the source was operated with ~ 6 mTorr (~ 0.8 Pa) of neutral argon gas flowing at 60 sccm and with a uniform magnetic field of ~ 900 Gauss. Helicon sources have windows of operation in the operational parameter space (neutral gas pressure and flow, external magnetic field and RF power), and this combination was found to give the helicon mode of operation reliably for the entire RF power scan. Since this was the first study in the PISCES-RF device, only parametric scans of the RF power were performed.

In Fig. 8, radial profiles of the plasma density (measured by the RF compensated Langmuir probe and assuming ions are singly-ionized Ar+, as shown previously using spectroscopy [56]) are shown as the forward RF power is increased from 2 to 10 kW. The density profiles are measured at the first port, 0.8 m away from the source (see Fig. 1). For these operating conditions, the plasma density at $r = 0$ peaks at ~ 6 kW of RF power input and then subsequently decreases. The highest plasma density at these conditions is $n \sim 2.5 \times 10^{19}$ m$^{-3}$. The plasma density profile measurements are extremely repeatable. Shown in Fig. 9 are the plasma densities measured at $r = 0$ cm, at port 1, located 0.8 m (in blue) and at port 2, located 1.5 m (in red) downstream of the source as a function of RF input power. For low RF powers, there is significant axial variation in the peak plasma density [1, 2, 4]. But at higher RF powers, the plasma density is more uniform along the source axis.

In the helicon mode, it is often observed that the peak density at one axial location initially increases with increasing RF power, but then saturates and even decreases with additional RF power (similar to what is shown in Fig. 9). This phenomenon is attributed to changes in the neutral depletion [57] near the center of the plasma due to strong ionization (fueling limitations) in the helicon mode and is related to density limits [58] in helicon sources.

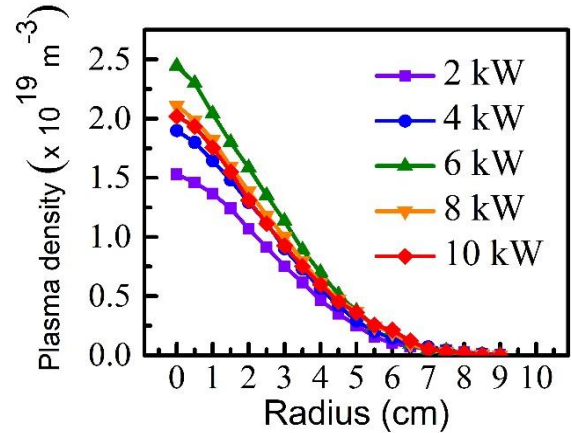

Fig. 8. Radial profiles of argon plasma density measured 0.8 m downstream of the RF helicon source.

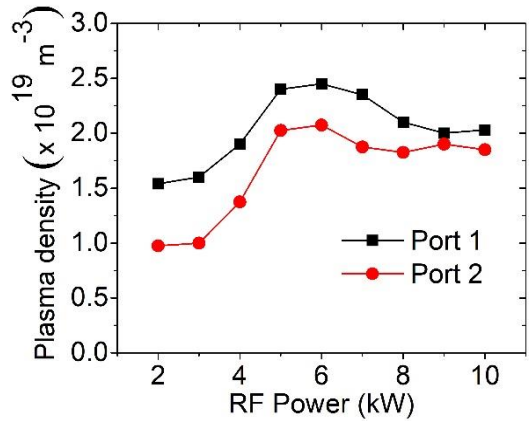

Fig. 9. Plasma electron density vs. RF power measured at $r = 0$ cm, at axially 0.8 m (black squares) and 1.5 m (red circles) downstream of the RF helicon source.

The radial profiles of the electron temperature, measured at the first port, 0.8 m away from the source (see Fig. 1), are shown in Fig. 10, as the forward RF power is increased from 2 to 10 kW. The results are repeatable on a shot-to-shot basis. However due to possible errors in fitting, the error estimated is ~ ± 0.5 eV. The electron temperature in this regime increases monotonically with increasing RF power. These electron temperature profiles are consistent with previous measurements in high-density helicon sources, albeit at lower RF powers [8, 24, 59]. The prominent electron temperature peak near the plasma edge is thought to be due to damping of the evanescent slow wave branch of the same dispersion relation that yields the classic helicon (whistler) wave [1 – 5, 59 – 62]. The only way to create an off-axis temperature peak in a cylindrically symmetric system is to heat the electrons off-axis, i.e., through damping of the slow wave on the electrons.

The electron temperatures measured at $r = 0$ cm at port 1, located 0.8 m (black squares) and the port 2, located 1.5 m (red circles) downstream of the RF helicon source are shown in Fig. 11. We find that in both locations, the electron temperature increases monotonically with RF power. For all RF powers, the electron temperature decreases with axial distance, consistent with previous work on helicons [1 – 5, 63].

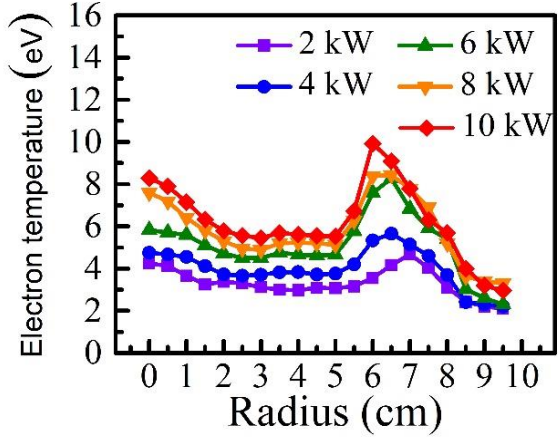

Fig. 10. Radial profiles of electron temperature measured 0.8 m downstream of the RF helicon source.

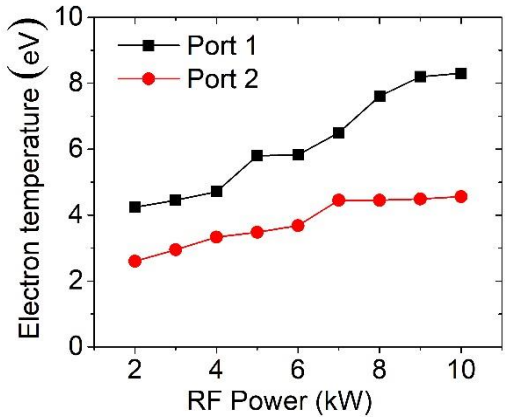

Fig. 11. Electron temperature vs. RF power measured at $r = 0$ cm, at axially 0.8 m (black squares) and 1.5 m (red circles) downstream of the RF helicon source.

**4.3 LIF measurements: RF power scan**

For the same operating conditions as described in section 4.2, an example of the raw argon ion LIF signal and the corresponding nonlinear fit are shown in Fig. 12. The fit yields an ion temperature of ~ 0.72 eV and a mean ion flow of ~ 440 m/sec. The laser was injected perpendicular to the magnetic field, thus providing a measure of the perpendicular bulk ion velocity and ion temperature. Careful observation of the plasma cross sections using 2-D fast imaging [56, 64] show that at high RF powers, the plasma center does not necessarily coincide with the chamber center. Typically, the difference between the plasma center and the chamber center is ~ 1.5 cm. Assuming this offset, these LIF measurements represent the perpendicular ion temperature and azimuthal ion velocity at $r = 1.5$ cm.

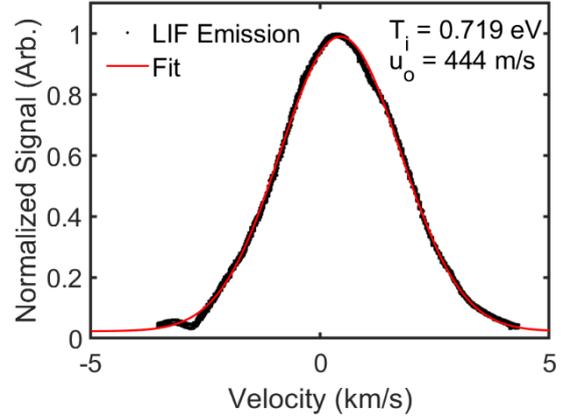

Fig. 12. Example of the LIF signal and the non-linear fit that yields an ion temperature of 0.72 eV and a mean azimuthal flow of ~ 440 m/s at $r = 1.5$ cm.

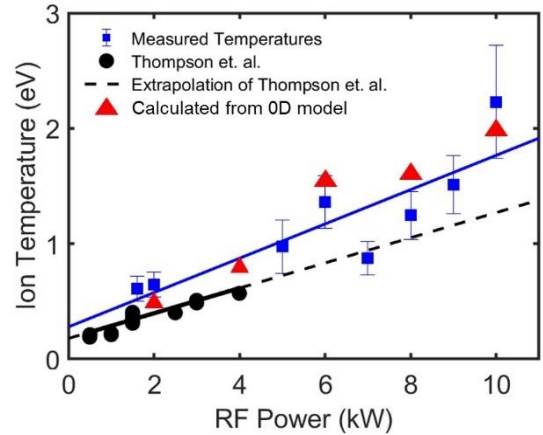

Fig. 13. Perpendicular ion temperature versus RF power for PISCES-RF (blue squares) and an extrapolation of previous measurements in RAID (black circles) [23]. Shown as red triangles are the calculated ion temperature values from a volume averaged 0-D model.

The perpendicular ion temperatures measured in the core plasma of PISCES-RF near the center ($r \sim 1.5$ cm) are shown in Fig. 13 (blue squares) as the RF power is increased from 1.5 to 10 kW. At 10 kW, we achieved an ion temperature of 2.2 eV. To our knowledge, these are the highest ion temperatures reported in a steady-state, high-density helicon plasma source. The measured ion temperatures are consistent with calculated ion temperature values (shown as red triangles in Fig. 13) using a simple volume averaged 0-D power balance model. In the model, it is assumed that electron-ion collisions are the dominant mechanism leading to ion heating, which is balanced by two ion heat loss mechanisms: convective parallel losses via the end of the machine and neutral heating due to ion-neutral collisions (see Appendix II for a brief description of the model).

In addition, an extrapolation to RF powers of 10 kW based on the previous study of the scaling of ion temperature with increasing RF power in RAID [23] is also shown (as black squares) in Figure 13. The power scaling is linear on both devices. The measured ion temperature in PISCES-RF is higher than the value predicted by the RAID scaling, most probably due to differences in the RF antenna design, gas feed and source parameters such as gas flow rate and neutral pressure. If the linear scaling of ion temperature with RF power holds even for much higher RF powers, it is anticipated that at RF powers > 100 kW (as expected in future PMI facilities), ion temperatures of more than 20 eV are achievable. Though the details of the source parameters might be different, these independent measurements and power scaling in similar sized cylindrical plasma devices suggest that divertor-like plasma conditions are achievable in high-power, steady-state helicon plasmas, similar to the MPEX device being developed at ORNL.

The bulk azimuthal ion velocity as a function of increasing RF power, measured at $r = 1.5$ cm is shown in Fig 14. We find that the mean azimuthal ion velocity increases slightly with increasing RF power. At certain RF powers (~ 5 kW), the plasma couples to a different equilibrium mode in PISCES-RF [65, 66] and the plasma center is displaced from the center of the chamber. During the transition between these two global equilibrium modes, the plasma is unstable with large, low-frequency oscillations in the order of a few Hz. Hence for RF powers of 3 kW and 4 kW, the source was too unstable for LIF measurements. The large error in the flow measurements at 5 kW in Fig. 14 is also due to the mode stability issue.

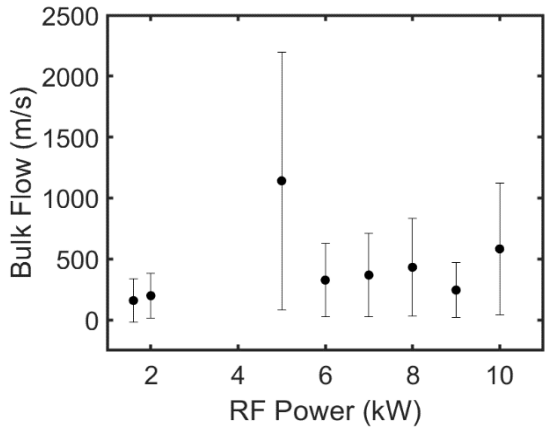

Fig. 14. LIF measurements of the mean azimuthal ion flow velocity versus RF power, measured at $r = 1.5$ cm.

## 5. Modeling of Neutrals in PISCES-RF

The fact that the plasma density does not increase linearly with RF power, as shown in Fig. 9, suggest that neutral depletion effects might be at play in determining the densities achievable in PISCES-RF. It has been shown in Proto-MPEX that understanding and controlling the neutral gas distribution is critically important for optimizing these high-power RF sources as PMI devices [34]. The location of the neutral gas injection, use of differential pumping and the use of skimmers to control excessive neutrals near the edge all have significant effects on plasma production. In these experiments, we specifically used end gas injection (Fig. 2) to replicate proto-MPEX operation and to provide input into the MPEX design.

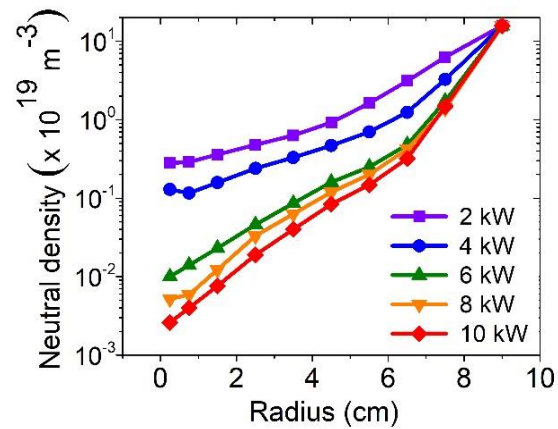

Fig. 15. Radial profiles of argon neutral density calculated from the 1-D Monte-Carlo simulations.

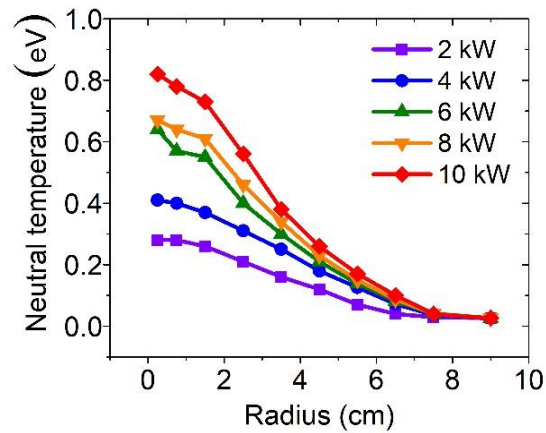

Fig. 16. Radial profiles of argon neutral temperature calculated from the Monte-Carlo simulations.

Because measurement of neutrals in high-density RF helicon plasmas is experimentally challenging [67 – 70], we used 1-D Monte Carlo simulations to provide insight into the neutral density and temperature using the measured radial profiles of plasma density, electron temperature, and ion temperature [71]. The simulation is constrained by the input gas pressure as measured with a neutral pressure gauge. Ionization by electron impact, neutral-neutral, and ion-neutral collisions are included in the model. Mean radial plasma and neutral flows and wall pumping are ignored. The radial profiles of the calculated neutral gas densities and the predicted neutral temperatures ($T_n$) are shown in Fig. 15 and Fig 16, respectively. For our measured plasma conditions, the model predicts strong neutral depletion at the center, consistent with previous studies of neutral depletion in helicon sources [57, 58]. The depth of the neutral depletion becomes stronger with increasing RF power. This also suggests that due to the severe neutral depletion, further increasing the RF power is not expected to yield any further increases in the plasma density, consistent with the data shown in Fig. 9.

The predicted neutral temperatures increase with increasing RF power. There are very few measurements of neutral temperatures in high-density helicon plasmas and all reported neutral temperatures are smaller than those predicted by this model. However, it is important to note that the previously reported neutral temperature measurements were all obtained at much lower RF powers than those used here [72].

The measured ion temperatures and calculated neutral temperatures from this study are normalized by the electron temperature and plotted in Fig. 17. We see that the normalized ion temperature (solid black squares) and the normalized neutral temperatures (solid red circles) increase linearly with RF power. That the normalized ion and neutral temperatures increase with RF power suggests that collisional processes, such as electron-ion, electron-neutral and ion-neutral collisions, play an ever-increasing role in the heating of the ions and neutrals at these higher plasma densities at the higher RF powers. The ions are heated to ~ 10 – 25% of the electron temperature while the neutral gas on axis is heated to a little less than half of the ion temperature. This suggests that further increasing in the ion temperature is possible with additional RF power input.

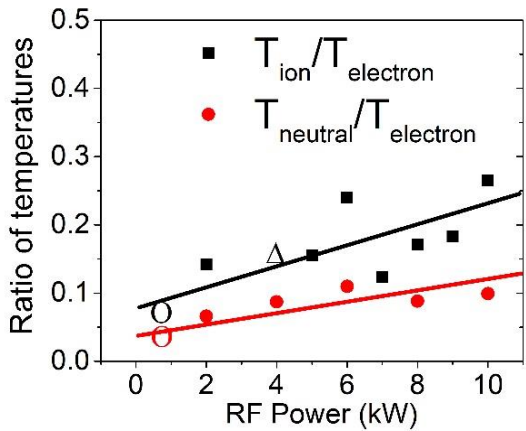

Fig. 17. Neutral temperature (red circle) and ion temperature (black square) normalized by the electron temperature versus RF power. The ion and electron temperatures are measured, and the neutral temperature is predicted from a 1-D Monte-Carlo simulations. The open triangle and circles are values of the same ratios as reported by other researchers [23, 72].

In Fig 17, we also check for consistency with previous measurements of ion, neutral and electron temperatures. The hollow black triangle and the hollow black circle are $T_i/T_e$ values from two previous studies [23, 72]. The hollow red circle is the measured $T_n/T_e$ from another study [72]. The measurements and modeling results reported here are thus consistent with previous studies in helicon sources at lower RF powers.

## 6. Summary and Discussions

These are the first measurements of plasma density, electron temperature, and ion temperature in steady state argon plasmas at RF powers as high as 10 kW. Such high power, steady-state source operation, on the order of several minutes, was made possible by using a novel DI-water-cooled RF antenna assembly with two concentric ceramic layers as the RF-transparent window.

The high density ($n > 10^{19}$ m$^{-3}$) helicon mode of operation is achieved ~ 1200 Watts of forward RF power for the magnetic fields > 800 G. Both the electron and the ion temperatures increase linearly with increasing RF power. At 10 kW, the electron temperature is > 8 eV and the ion temperature is > 2 eV, at plasma densities of ~ 2 x 10$^{19}$ m$^{-3}$ at ~ 1 m away from the RF plasma source.

In Ref 12, Table 3 gives a comparison of the plasma parameters of currently active PMI devices (mostly reflex arc and cascaded arc sources) to the ITER divertor conditions for the output to input power amplification ratio using the deuterium tritium nuclear fusion reactions ($Q_{DT}$) = 10 scenario. Table 5 of the same Ref. 12 gives the expected plasma parameters achievable for planned linear PMI devices. At 10 kW of RF power in PISCES-RF, we achieve similar conditions (plasma density, electron temperature, ion temperature, ion flux etc.) in argon plasma.

With 20 kW of RF power, the same source in PISCES-RF produces hydrogen and helium plasma densities and particle fluxes (hydrogen plasma density > 10$^{19}$ m$^{-3}$ and ion flux > 2 x 10$^{23}$ m$^{-2}$sec$^{-1}$ even 1.5 m away from the RF source) [25], similar to those achieved by other linear PMI devices trying to replicate divertor-like plasma conditions [12]. The capability of steady state operations at these elevated densities and particle fluxes would allow delivery of record plasma fluences on material targets placed in PISCES-RF.

The 1-D Monte-Carlo simulations of the neutral densities as well as the measurements of the electron density as a function of increasing RF power suggest that at high powers, the core of the plasma becomes starved of neutrals, which in turn prevents achieving higher plasma densities. To obtain higher density plasma production in similar devices, new gas injection schemes will be required, such as injection of the gas near the core of the plasma or cryogenic pellet fueling, as opposed to gas puffing at the edge.

Finally, we note that while the perpendicular ion temperature increases linearly with increasing RF power, previous studies have shown that the parallel ion temperature in helicon sources is nearly independent of RF power [23] and depends weakly on the magnetic field [16 – 18, 51]. Since the perpendicular ion temperature in these measurements increases linearly with both the RF power and the magnetic field [16 – 18], high-power RF plasmas at high magnetic fields might facilitate studies with even larger ion temperature anisotropy than were previously possible, which is of interest to the space plasma community for laboratory modeling of the physics of space plasma phenomena.

## APPENDIX I: Numerical corrections for broadening due to oversampling during LIF measurements

Care must be taken when evaluating the output signal of the lock-in amplifier. Particularly, the ratio of the total laser scan time ($T_{scan}$) and the lock-in time constant ($RC$) is an important factor in avoiding systematic errors that would otherwise lead to artificial broadening of the VDF. The effect of changing the values of $T_{scan}/RC$ on the artificial broadening of an undistorted signal (in black) is shown in Fig. 5. When the total laser scan time is not much larger than the lock-in time constant ($T_{scan}/RC = 2$ shown in blue and $T_{scan}/RC = 10$ shown in red), the width is artificially exaggerated, and the peak shifts. This effect is similar to applying a low-pass filter with a sharp cutoff to the input signal. The distortion is small when the total laser scan time is much larger than the lock-in time constant ($T_{scan}/RC = 30$, shown in purple and $T_{scan}/RC = 100$, shown in green). For LIF measurements at the highest RF powers in this study, this over-integration effect was non-negligible.

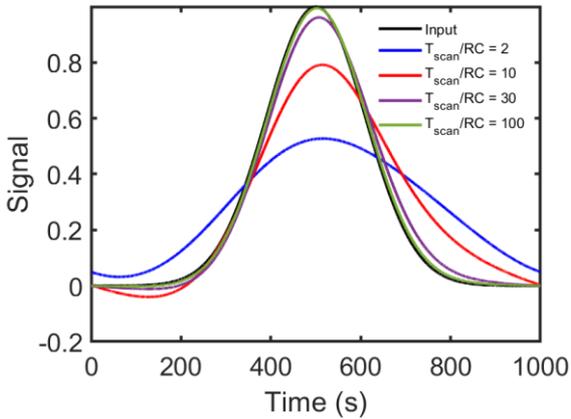

Fig. 18. Artificial line broadening of the LIF signal due to a systematic error while using low values of $T_{scan}/RC$.

Mathematically, the output of the lock-in amplifier is a convolution of the input signal $f(t)$ and the reference signal from the modulator $g(t)$. A time-space convolution is a product in Fourier space, given as

$$f(t) * g(t) \xrightarrow{FT} F(\omega)G(\omega), \qquad (2)$$

where $F(\omega)$ and $G(\omega)$ are the corresponding Fourier transforms. The convolution is then low-pass filtered, i.e., integrated over the time constant. Ideally, the smoothing effects of the low-pass filter are removeable by dividing the convolution by the transfer function of the low-pass filter in frequency space and then inverse fast Fourier transforming the result back into time space. In practice, however, the zero amplitude of an ideal low-pass filter transfer function at high frequencies leads to an artificial amplification of any high frequency noise present in the signal. Hence, the division by the filter transfer function in frequency space is only performed for a limited range of frequencies. A careful numerical study was performed to determine the optimal fraction of the low-pass filter transfer function needed to remove over-integration effects. Independent of the value of $T_{scan}/RC$, only the leading 0.2% of the filter was necessary to reconstruct the original time series. Any more than 0.2% resulted in negligible differences in the reconstructed time series and introduced significant high-frequency noise. The result of the analysis is shown in Fig. 6. We see that the deconvolution technique yields the original signal, independent of the value of $T_{scan}/RC$.

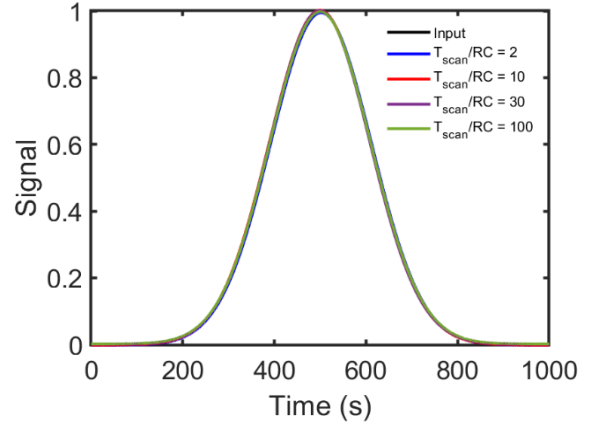

Fig. 19. Deconvolution techniques can correct the artificial line broadening of the LIF signal due to low values of $T_{scan}/RC$.

## APPENDIX II: Volume averaged 0-D power balance model to estimate ion temperature

Here we describe a simple volume averaged 0-D power balance model [73] to estimate the ion temperature from the experimentally measured electron density and temperature profiles and numerically estimated neutral densities. The model was developed to help identify likely ion heating and loss mechanisms and also to provide rough estimates of the expected ion temperatures. We assume collisional heating of ions via electron-ion collisions. In equilibrium, the heating is balanced by two loss terms, the parallel loss of ions due to convective transport at the end of the cylindrical device and the heating of the neutrals by the ions through ion-neutral collisions. Thus, we have the following equation:

$$\frac{(T_e - T_i)}{\tau_{eq}} = \frac{T_i}{\tau_{\|}} + \frac{(T_i - T_n)}{\tau_{in}}, \qquad (3)$$

where $T_e$, $T_i$ and $T_n$ are the electron, ion and neutral temperatures, respectively and $\tau_{eq}$, $\tau_{\|}$ and $\tau_{in}$ are the characteristic time scales of electron-ion equilibration, parallel end losses, and ion-neutral collisions. Consistent with previous studies where both the ion and the neutral argon temperatures were measured [72] and the numerical studies described in section 5, we assume $T_n = T_i/2$ for simplification. We use standard expressions for the electron-ion collision frequencies and the ion-neutral collision frequencies from the plasma formulary. For the parallel loss time scales, we assume $\tau_{||} = \frac{L_{||}}{C_s \cdot M}$, where $L_{||}$ is the length of the device, $C_s$ is the ion sound speed and $M$ is the Mach number. Using typical values for argon plasma in CSDX [8, 73], we have

$$\tau_{in} = (3.7 \times 10^{-5}) \left(\frac{10^{19}}{n_n}\right)\left(\frac{1}{\sqrt{T_i}}\right), \quad (4)$$

$$\tau_{eq} = (1.3 \times 10^{-4}) \left(\frac{10^{19}}{n_i}\right)\left(\sqrt{T_e}\right)^3, \quad (5)$$

and

$$\tau_{||} = (6.2 \times 10^{-3}) \left(\frac{1}{\sqrt{T_e}}\right), \quad (6)$$

where the temperatures are in eV and densities are in m$^{-3}$. Using the volume averaged experimental values of the electron temperature and density and the numerical value of the neutral plasma density, we numerically solve Eqn. (3) to estimate the values of $T_i$.

We note that the results from this simple volume averaged 0-D model are consistent with the experimentally measured ion temperatures using LIF when we take the radial extent of the volume of integration to be $r \sim 6$ cm. This is most probably because the high density helicon plasma column is well defined for $r < 6$ cm, as seen from the experimentally measured radial profiles of plasma density (Fig. 8) and the numerically calculated radial profiles of the neutral density (Fig. 15). The edge plasma for $r > 6$ cm is weakly ionized and hence has very different plasma characteristics. Beyond $r > 6$ cm, the simple 0-D volume integrated model breaks down. We also note that the experimental measurements using LIF were performed only near the high-density helicon core ($r \sim 1.5$ cm).

**Acknowledgments**

This research is sponsored by the Office of Fusion Energy Sciences, U.S. Department of Energy, under contract DE-FG02-07ER54912. M.P. and E.E.S. were supported by NSF grant. PHY-1902111 and Department of Energy Grant DE-SC0020294. S.C.T. thanks Mr. Leopoldo Chousal for his help in designing and building the DI-water cooled RF source assembly.